# Source Number Estimation via Entropy Estimation of Eigenvalues (EEE) in Gaussian and Non-Gaussian Noise

Hamid Asadi, and Babak Seyfe, *Senior Member, IEEE*

*Abstract*— In this paper, a novel method based on the entropy estimation of the observation space eigenvalues is proposed to estimate the number of the sources in Gaussian and Non-Gaussian noise. In this method, the eigenvalues of correlation matrix of the observation space will be divided by two sets: eigenvalues of signal subspace and eigenvalues of noise subspace. We will use estimated entropy of eigenvalues to determine the number of sources. In this method we do not need any a priory information about signals and noise. The advantages of the proposed algorithm based on the performance is compared with the existing methods in the presence of Gaussian and Non-Gaussian noise. We have shown that our proposed method outperforms those methods in the literature, for different values of observation time and Signal to Noise Ratio, i. e. SNR. It is shown that the algorithm is consistent and also its probability of false alarm and probability of missed detection tend to zero for long observation time.

## I. INTRODUCTION

Signal enumeration has an important role in several fields such as brain imaging [1], neural networks [2], audio signals separation [3] and finance [4]. Also, the most important problem in sensor array signal

The authors are with Information Theoretic Learning System Laboratory (ITLSL), at the Department of Electrical Engineering, Shahed University, Tehran, Iran. The corresponding author is Babak Seyfe (e-mails: {h.asadi, seyfe} @shahed.ac.ir).

processing is to estimate the coordinates of the sources that are emitting signals. Since the azimuth and elevation angles of each point in three dimensional spaces are estimated by the measurement of direction of arrival (DOA) by an array of sensors, the number of sources must be known [5].

Some of the proposed solutions for array signal processing case are based on information theoretic criteria. *Akaike-Information Criterion (*AIC*)* [6], *Minimum Description Length (*MDL*)* [7], *Bayesian Information Criterion (*BIC*)* [8] and *Predictive Description Length (*PDL*)* [9] can be considered as four such ways in this approach. AIC and MDL minimize the Kullback-Leibler distance between the observations and the data model that is estimated by maximum likelihood estimator. MDL estimator fails to estimate number of sources in low SNR or small number of observation samples [10] but it is consistent and for large observation times, the probability of error that is equal to the probability of *under estimation* tends to zero, approximately [11]. While AIC estimator has a better performance in low SNR regime and small number of observation samples [10], this estimator is not consistent in the same conditions and its error probability or probability of *overestimation* does not tend to zero [11]. The proposed PDL method uses the projection of sample covariance matrix on to the signal and noise subspaces to estimate the number of sources based on the maximum likelihood (ML) estimate of the past data and at each time, new data will be updated by a recursive algorithm [9].

In [12], the probability of missed detection in MDL method is derived using the Tracy-Widom distribution for the largest eigenvalues and Gaussian distribution by updating a threshold using a first order Taylor series. Also it is shown that the statistical performance of the MDL is approximately the same under deterministic signal models [12]. In [13], new frameworks for analytically evaluating the statistical performance of eigen decomposition based detectors are considered. Also, in there, the exact and approximate asymptotic bounds of the overestimation probabilities of AIC and MDL are discussed. The number of sources is approximated by Fishler, Grosmann and Messer [14] via considering the case of both Gaussian sources and Gaussian noise. They use the asymptotic distribution of the eigenvalues of the empirical correlation matrix and it is shown that the performance of MDL estimator is not very sensitive to the actual distribution of source signals. In [15] the problem of optimal rank estimation is considered by developing a decision-theoretic rank estimation such as a min-max algorithm. In [16], the number of sources is estimated using hypothesis testing as a Neyman-Pearson method and at each step testing the significance of eigenvalues as arising from a signal will be evaluated. Also, the probability of overestimation and its bounds are computed. In [17], a nonparametric method that depends on choosing a function of estimated eigenvalues from sample covariance matrix that is proposed to enumerate the number of sources. In [18], the number of sources is estimated based on the correlation matrix decomposition method and it uses the null space of correlation matrix to estimate the directions of arrival of the sources. In [19], the problem of signal enumeration is considered in an information theoretic criteria framework. In there, probability density function of eigenvalues is estimated based on the observation samples and used to signal enumeration in addition to the probability density function of observation. In [20], signal enumeration is investigated via bootstrap method to estimate the source number in an environment with a Gaussian noise mixture. Also, in [21] signal enumeration problem using a criterion based on generalized Bayesian information Criterion, i.e. GBIC, is investigated.

In the most published works [6-19, 21], the authors assumed that the noise is Gaussian and white both spatially and temporally. But in many cases due to impulsive nature of noise, such as sonar [22-23] and impulsive man-made noise in urban and indoor wireless systems [24-25], the noise model is often non-Gaussian or impulsive. Based on the most knowledge of the authors of this paper, the only work that is assumed non-Gaussian noise in the literature is [20].

In this paper, we present and analyze a new estimation algorithm in Gaussian and non-Gaussian noise based on the information theoretic criteria using the concept of amount of information earned from eigenvalues of the empirical correlation matrix of the observations. In our proposed method, called *Entropy Estimation of Eigenvalues (EEE)*, we do not need to know the probability distribution or any a priori information about the observations. The difference between two sets of eigenvalues (noise subspace eigenvalues and signal subspace eigenvalues) can be considered by using the entropy as a measure of uncertainty. We only assume that the sources are non-coherent, which is an advantage in comparison with other methods using additional assumptions [15], [16]. The entropy measures the uncertainty of random variables or amount of information of them without using any additive assumption [26]. We have shown that the proposed method is comparable and better than the best methods reported in literature in the presence of Gaussian noise and it outperforms them in the presence of Non-Gaussian noise drastically. Also, it is shown that the algorithm has an outstanding accuracy in limited number of observations compared with other methods.

This paper is organized as followings: in section II, we present models and preliminaries. In section III, we present our motivation to signal enumeration and an estimator to entropy estimation. Also, the entropy estimation of eigenvalues will be demonstrated as a rank estimator in this section. Sections IV and V, consist of the performance analysis of the proposed method, and simulation results, respectively. Finally, section VI concludes the paper.

*Notation:* In this paper, we use matrix and vectors by uppercase and lowercase, bold, respectively. The form $(\hat{\cdot})$ means the estimated unknown parameters. Also $\mathcal{N}(\mathbf{0}, \mathbf{\Lambda})$ denotes the Gaussian distribution with zero vector mean and positive definite covariance matrix $\mathbf{\Lambda}$. $\hat{H}(\lambda)_i^j$ is the estimated Shannon entropy of random variable $\lambda$ based on the observed samples $\left( \hat{\lambda}_i, ..., \hat{\lambda}_j \right)$.

## II. SIGNAL SUBSPACE MODEL

### A. Problem Formulation

Assume that $K$ sources send their signals, independently, to $P$ ($P>K$) sensors at the receiver. The observed signal at the receiver in time $t_i$ is denoted by

$$\mathbf{x}(t_i) = \mathbf{A}\mathbf{s}(t_i) + \mathbf{n}(t_i), \quad 1 \le i \le N, \tag{1}$$

where $\mathbf{x}(t_i)$ and $\mathbf{A}$ are the observed vector with the dimension of $P$ in time $t_i$ and steering matrix with the size of $P \times K$, respectively. Also the components of the vector $\mathbf{s}(t_i) = [s_1(t_i), \ldots, s_K(t_i)]^T$ are white, real, stationary processes with zero mean and positive definite covariance matrix $\mathbf{R}_S = diag(l_{S,1}, \ldots, l_{S,K})$ where $l_{S,i}$, $1 \leq i \leq K$ is the received power from the $i^{th}$ source. The noise samples are assumed to be additive white zero mean (Gaussian or non-Gaussian type) with unknown noise power $(\sigma^2)$ which are independent from $\mathbf{s}(t_i)$. Noting (1), the population covariance matrix of the received signal from eigen-decomposition can be formulated as follows [7]

$$
\begin{aligned}
\mathbf{C} &= E\left[\mathbf{x}(t_i)\mathbf{x}^H(t_i)\right] \\
&= \mathbf{A}^H \mathbf{R}_S \mathbf{A} + \sigma^2 \mathbf{I}_P \\
&= \mathbf{U}\boldsymbol{\Lambda}\mathbf{U}^H,
\end{aligned}
\tag{12}
$$

where in (2), $\mathbf{I}_P$ denotes the $P$-dimensional identity matrix, superscript $H$ shows the Hermitian of matrix, $\boldsymbol{\Lambda} = diag(\lambda_1, \lambda_2, \ldots, \lambda_P)$ is the eigenvalue matrix of the observation space and $\mathbf{U}$ is the eigenvector matrix of the observations with the size of $K \times P$. Equation (2) shows that $\lambda_1 > \lambda_2 > \ldots > \lambda_P$ denotes the eigenvalues of the observation space and they are ordered in descending order of magnitudes [10]. It is shown that the first $K$ eigenvalues, i.e. $\lambda_1, \lambda_2, \ldots, \lambda_K$, belongs to the signal subspace and the other smallest $(P-K)$ eigenvalues assumed as the eigenvalues of the noise subspace which are equal to $\sigma^2$ [21].

In a limited observation time, we can approximate the sample covariance matrix of $\mathbf{x}(.)$ at $N$ observation times, by

$$
\begin{aligned}
\hat{\mathbf{C}} &= \frac{1}{N}\sum_{i=1}^{N}\mathbf{x}(t_i)\mathbf{x}^H(t_i) \\
&= \hat{\mathbf{U}}\hat{\boldsymbol{\Lambda}}\hat{\mathbf{U}}^H, \quad \hat{\boldsymbol{\Lambda}} = diag\left(\hat{\lambda}_1, \ldots, \hat{\lambda}_P\right).
\end{aligned}
\tag{23}
$$

By the weak law of large numbers (WLLN), $\hat{\mathbf{c}}$ and $\hat{\boldsymbol{\Lambda}}$ in (3) tend to the covariance matrix $\mathbf{C}$, and eigenvalue matrix, $\boldsymbol{\Lambda}$ asymptotically as $N \to \infty$, respectively [27].

Now we define a new variable $\lambda$ with the observed samples of $\left(\hat{\lambda}_1, \ldots, \hat{\lambda}_P\right)$. We would like to bring the attentions on that the $\lambda$ is different from *N*-observation time to another *N*-observation time. Then we can assume that $\lambda$ is a random variable. We consider this randomness via entropy estimation as criterion to estimate the number of independent sources.

### III.   DETECTING THE NUMBER OF SIGNALS BASED ON ENTROPY ESTIMATION

#### A.   Motivation: uncertainty or amount of information earned from estimated eigenvalues

We draw the attentions on that $\hat{\mathbf{\Lambda}}$ is a random matrix in the limited time of observation ($N < \infty$). This allows us to use entropy of eigenvalues for our algorithm to enumerate the number of the source signals. To measure the uncertainty or amount of information that can be earned from eigenvalues, we use a criterion based on information theoretic measures such as entropy [27]. Based on entropy properties, we can see that the amount of information of eigenvalues vector equals to only first ($K+1$) eigenvalues, because

$$\left( \lambda_1, ..., \lambda_P \right) = \left( \lambda_1, ..., \lambda_K, \underbrace{\sigma^2, ..., \sigma^2}_{P-K} \right),$$

(4)

and it is seen that the last (*P-K*) eigenvalues equal to $\sigma^2$ and based on entropy properties, the amount of information earned from one $\sigma^2$, equals to the amount of information observed from multiple $\sigma^2$, so we can see that

$$\hat{H}\left( \lambda \right)_{K+1}^{P} = \hat{H}\left( \lambda \right)_{K+2}^{P}$$
$$\vdots$$
$$= \hat{H}\left( \lambda \right)_{P}^{P},$$

(5)

or

$$\hat{H}\left( \lambda \right)_{1}^{K+1} = \hat{H}\left( \lambda \right)_{1}^{K+2}$$
$$\vdots$$
$$= \hat{H}\left( \lambda \right)_{1}^{P}.$$

(6)

Equations (5) and (6) will be our motivation to enumerate the number of signals impinging to a sensor array. So we need to estimate the entropy of eigenvalues.

## B. Entropy Estimation using Kernel Function

Because of our above motivations, we need to estimate the entropy of the eigenvalues based on the observations. In [30], Paninski showed that the estimation of probability spaces, which are needed for an entropy estimator, is an important challenge and could be a much difficult problem, and the solution to solve this problem is: *nonparametric estimators*. Also, three algorithms were investigated by Paninski to estimate the entropy: maximum likelihood (MLE) [31-32], Miller-Madow bias correction [33], and jackknifed-MLE [34]. All of the above investigated algorithms are nonparametric estimators that it means they need to estimate the probability space of measured under consideration data. We devise an estimator that estimate the entropy directly from data without any need to know about its probability distribution function. This method is based on the using of the kernels which are developed by Parzen and Rosenblatt [35-36]. Kernels are used for example in regression for estimation of random variables when conditioned, or in probability density estimation to estimate the PDF of random variables [37].

Based on the difficulties of the density estimation to estimate the entropy in high dimensions and with few samples, the kernel functions [37], will be useful functions to estimate entropy directly from samples without any need to PDF estimation. Kernel-based entropy estimator from eigenvalues, will be assumed with the following properties [34]

$$\mathcal{K}\left(x\right) \geq 0, \tag{7}$$

$$\int_{R} \mathcal{K}\left(x\right) dx = 1, \tag{8}$$

$$\lim_{x \to \infty} \left| x\, \mathcal{K}\left(x\right) \right| = 0, \tag{9}$$

where the kernel $\mathcal{K}(x)$ is a symmetric function. A scaled kernel function, $\mathcal{K}_h(x)$ with smoothing parameter $h > 0$ that is called *bandwidth*, can be defined as

$$\mathcal{K}_h\left(x\right) = \frac{1}{h}\mathcal{K}\left(\frac{x}{h}\right). \tag{10}$$

Smoothing parameter, window width or bandwidth is a free parameter that affects the estimation and estimation of the bandwidth for a kernel is considered in some articles [41-43].

Some of the popular kernels are listed in [37, 40]. It can be readily seen that for most of listed kernels in [40] and all of the kernels listed in [37], $\mathcal{K}(x) = 0$ for $|x| > 1$.

To estimate the $\hat{H}\left(\lambda\right)_i^j$, the following equation will be proposed as a (kernel) entropy estimator [37]

$$\hat{H}\left(\lambda\right)_i^j = \left(\frac{-1}{j-i+1}\right)\sum_{k=i}^{j}\log\left(\frac{1}{j-i+1}\sum_{l=i}^{j}\mathcal{K}_h\left(\hat{\lambda}_k - \hat{\lambda}_l\right)\right). \tag{11}$$

In [37], it is shown that the equation (11) can be used as the Shannon entropy of the $\lambda$ based on $\left(\hat{\lambda}_i, ..., \hat{\lambda}_j\right)$ samples of observations.

C.  *Entropy Based estimation of the Number of Sources*

We define

$$F\left(i\right) \triangleq \hat{H}\left(\lambda\right)_i^P \tag{12}$$

We consider $F(i)$ in the joint limit $P, N \rightarrow \infty$ with $P/N$ fixed and show that it is a concave ($\cap$) function of $i$ for $i \leq K+1$ (Here, concavity means that the second order differentiation of $F(i)$ as $P, N \rightarrow \infty$ with $P/N = c > 0$, that is discrete in time, is negative for all values of $i \leq K+1$).

**Lemma 1**: *For large values of number of sensors and observation time, i.e., $P, N \rightarrow \infty$, $F(i)$ is a concave function ($\cap$) of i for $i \leq K+1$.*

**Proof**:  Define the first order difference of $F(i)$ as follows

$$\begin{aligned}\Delta F\left(i\right) &= F\left(i+1\right) - F\left(i\right)\\ &= \hat{H}\left(\lambda\right)_{i+1}^P - \hat{H}\left(\lambda\right)_i^P. \end{aligned} \tag{13}$$

Based on the definition of kernel functions,

$$\mathcal{K}_h\left(\hat{\lambda}_j - \hat{\lambda}_k\right) = (1/h)\,\mathcal{K}\left(\left(\hat{\lambda}_j - \hat{\lambda}_k\right)\big/ h\right), \tag{14}$$

$F(i)$ can be written as the following

$$\begin{aligned}
F(i) &= \hat{H}\left(\lambda\right)_i^P \\
&= \left(\frac{-1}{P-i+1}\right)\sum_{j=i}^{P}\log\left(\frac{1}{P-i+1}\sum_{k=i}^{P}\mathcal{K}_h\left(\hat{\lambda}_j - \hat{\lambda}_k\right)\right) \\
&= \left(\frac{-1}{P-i+1}\right)\sum_{j=i}^{P}\log\left(\frac{1}{h(P-i+1)}\left(\mathcal{K}(0) + \sum_{\substack{k=i \\ k\neq j}}^{P}\mathcal{K}\left(\frac{\hat{\lambda}_j - \hat{\lambda}_k}{h}\right)\right)\right).
\end{aligned} \tag{15}$$

After some simplifications, the above equation can be written as

$$F(i) = \log\left(P-i+1\right) + \log\left(h\right) + \left(\frac{-1}{P-i+1}\right)\sum_{j=i}^{P}\log\left(\mathcal{K}(0) + \sum_{\substack{k=i \\ k\neq j}}^{P}\mathcal{K}\left(\frac{\hat{\lambda}_j - \hat{\lambda}_k}{h}\right)\right). \tag{16}$$

So $\Delta F(i)$ is as following

$$\begin{aligned}
\Delta F(i) &= \log\left(\frac{P-i}{P-i+1}\right) + \left(\frac{-1}{P-i}\right)\sum_{j=i}^{P}\log\left(\mathcal{K}(0) + \sum_{\substack{k=i \\ k\neq j}}^{P}\mathcal{K}\left(\frac{\hat{\lambda}_j - \hat{\lambda}_k}{h}\right)\right) \\
&\quad - \left(\frac{-1}{P-i+1}\right)\sum_{j=i+1}^{P}\log\left(\mathcal{K}(0) + \sum_{\substack{k=i \\ k\neq j}}^{P}\mathcal{K}\left(\frac{\hat{\lambda}_j - \hat{\lambda}_k}{h}\right)\right).
\end{aligned} \tag{17}$$

In [44], for optimum values of the smoothing parameter, i.e. the bandwidth $h$, it is mention that "*the ideal window width will converge to zero as the sample size increases*" that it means $h \to 0$ when $P, N \to \infty$. Also The first $(K+1)$ samples of $\lambda$ are different from each other, and then for large values of observations, i.e. $N$, and based on the equation (9), we have

$$\lim_{h \to 0} \mathcal{K}\left(\frac{\hat{\lambda}_j - \hat{\lambda}_k}{h}\right) = 0 \quad for \ \ j \neq k \ \ and \ \ j, k \leq (K+1). \tag{18}$$

So we have

$$\lim_{N \to \infty} \Delta F\left(i\right) = \lim_{P \to \infty} \log\left(\frac{P-i}{P-(i-1)}\right), \ \ i \leq K+1. \tag{19}$$

It is seen that the first order differentiation of $F(i)$, e.g. $\Delta F(i)$, for $i \leq K+1$ is a negative decreasing function of $i$, asymptotically, and then the second differentiation of $F(i)$ is negative for all values of $i \leq K+1$, then it is a concave function ($\cap$) in this area.

<div align="right">■</div>

**_Theorem 1_**: *The number of independent sources (noise excluded) in a sensor array can be estimated as follows for large number of observations, i.e.* $N \to \infty$

$$\forall 1 \leq i \leq P-1 : \hat{K} = \arg\min_i \left(\hat{H}\left(\lambda\right)_{i+1}^{P} - \hat{H}\left(\lambda\right)_i^{P}\right), \tag{20}$$

*or*

$$\forall 1 \leq i \leq P-1 : \hat{K} = \arg\max_i \left(\hat{H}\left(\lambda\right)_1^{i+1} - \hat{H}\left(\lambda\right)_1^{i}\right), \tag{21}$$

*where* $\hat{H}\left(.\right)$ *denotes the estimation of Shannon Entropy.*

**_Note_**: In the theorem 1, (20) and (21) are originated from equations (5) and (6), respectively. In the following, we prove (20) and the prove procedure of (21) is the same as (20), and then we neglect it.

**_Proof of Theorem 1_**: As it is proved for lemma 1, $F(i)$ is a concave function ($\cap$) of $i$ for $i \leq K+1$. So for large values of observation time, $N$, we have

$$i \leq K+1 : \quad \Delta F\left(i\right) < 0, \quad and \quad F\left(i+1\right) < F\left(i\right). \tag{22}$$

For $K+1 \leq i \leq P$, $\hat{\lambda}_i$'s equal to $\hat{\sigma}^2$ and then we can write

$$\lim_{N\to\infty}\Delta F(i)=\lim_{N\to\infty}\left[\left(\frac{-1}{P-i}\right)\sum_{j=i+1}^{P}\log\left(\frac{1}{P-i}\sum_{k=i+1}^{P}\mathcal{K}_h\left(\hat{\lambda}_j-\hat{\lambda}_k\right)\right)-\left(\frac{-1}{P-i+1}\right)\sum_{j=i}^{P}\log\left(\frac{1}{P-i+1}\sum_{k=i}^{P}\mathcal{K}_h\left(\hat{\lambda}_j-\hat{\lambda}_k\right)\right)\right]$$

$$=\left(\frac{-1}{P-i}\right)\sum_{j=i+1}^{P}\log\left(\mathcal{K}_h(0)\right)-\left(\frac{-1}{P-i+1}\right)\sum_{j=i}^{P}\log\left(\mathcal{K}_h(0)\right) \tag{23}$$

$$=\log\left(\mathcal{K}_h(0)\right)-\log\left(\mathcal{K}_h(0)\right)$$

$$=0,\qquad\qquad K+1\ \le i\le P$$

The equations (22) and (23), mean that $F(i)$ for $i\le K+1$ is a decreasing function of $i$, and for $i\ge K+1$ is a non-decreasing function of $i$, respectively, and then we can conclude that $F(i)$ reaches to its minimum at $i=K+1$.

For $N\to\infty$, $F(i)$ is a concave function for $i\le K+1$ and has a constant values for $i\ge K+1$, then we can say that $\Delta F(i)$ has only one minima at $i=K$ and then the following criterion can be used to determine the number of sources.

$$\forall 1\le i\le P-1:\hat{K}=\arg\min_i\left(\hat{H}\left(\lambda\right)_{i+1}^{P}-\hat{H}\left(\lambda\right)_i^{P}\right). \tag{24}$$

∎

Based on equation (11), (24) can be rewritten as

$$\forall\ 1\le i\le P-1:$$

$$\hat{K}=\arg\min_i\left(\left(\frac{-1}{P-i}\right)\sum_{j=i+1}^{P}\log\left(\frac{1}{P-i}\sum_{k=i+1}^{P}\mathcal{K}_h\left(\hat{\lambda}_j-\hat{\lambda}_k\right)\right)-\left(\frac{-1}{P-i+1}\right)\sum_{j=i}^{P}\log\left(\frac{1}{P-i+1}\sum_{k=i}^{P}\mathcal{K}_h\left(\hat{\lambda}_j-\hat{\lambda}_k\right)\right)\right).$$

$$\tag{25}$$

We call this criterion as *Entropy Estimation of Eigenvalues* (*EEE*). The performance analysis of EEE (probability of missed detection and probability of false alarm) will be considered in the next section.

## IV. Performance Analysis of EEE

In this section, we show the performance of proposed algorithm, EEE, analytically when $N$ tends to infinity for equation (20). Because the performance of EEE algorithm based on (20) and (21) are the same, we analyze the performance of the proposed algorithm based on (20). This section consists of two parts: probability of false alarm and probability of missed detection.

### A. Probability of false alarm(over estimation)

In this subsection, the false alarm (over estimation) probability of the proposed algorithm will be discussed. For the start, we rewrite the criterion for signal enumeration that is

$$\hat{K} = \arg \min_{1 \leq i \leq P-1} \left( \hat{H}(\lambda)_{i+1}^{P} - \hat{H}(\lambda)_{i}^{P} \right)$$
$$= \arg \min_{1 \leq i \leq P-1} \left( \Delta F(i) \right). \tag{26}$$

For true number of sources $K$, the overestimation occurs when

$$\Delta F(K+n) < \Delta F(K-m), \forall\ 0 \leq m \leq K-1, \forall\ 1 \leq n \leq P-K, \tag{27}$$

and then the probability of false alarm will be

$$P_{fa} = P\left( \Delta F(K+n) < \Delta F(K-m) \right), \forall\ 0 \leq m \leq K-1, \forall\ 1 \leq n \leq P-K. \tag{28}$$

Based on equation (23), $\lim_{N \to \infty} \Delta F(K+n) = 0$, and we can write

$$\lim_{N \to \infty} P_{fa} = \lim_{N \to \infty} P\left( 0 < \Delta F(K-m) \right), \forall\ 0 \leq m \leq K-1. \tag{29}$$

Based on equation (19), we have

$$\lim_{N \to \infty} P_{fa} = \lim_{N \to \infty} P\left( 0 < \log\left( \frac{P-(K-m)}{P-(K-m)+1} \right) \right), \forall\ 0 \leq m \leq K-1. \tag{30}$$

But it is seen that $\left( (P-(K-m)) / (P-(K-m)+1) \right)$ is always smaller than 1, and then

$$\lim_{N \to \infty} P_{fa} = 0. \tag{31}$$

**Note 1**: For criterion given in (21), false alarm occurs when $\Delta F\left(K+n\right) > \Delta F\left(K-m\right)$, for $0 \le m \le K-1$, and $1 \le n \le P-K$. For computation of the false alarm probability of this criterion, we should obey a same procedure as it is mentioned for (20).

### B. Probability of Missed Detection

Missed detection occurs when $\hat{K} < K$. It means that this event happens when $\Delta F\left(K-k\right) < \Delta F\left(K+l\right)$, for $1 \le k \le K-1$, and $0 \le l \le P-K$. Based on [11], we will assume $l = 0$. If we define $P_{missed}$ as missed detection probability, then

$$
\begin{aligned}
P_{missed} &= P\left(\Delta F\left(K-k\right) < \Delta F\left(K\right)\right) \\
&= P\left(\Delta F\left(K-k\right) - \Delta F\left(K\right) < 0\right), \forall 1 \le k \le K-1.
\end{aligned}
\tag{32}
$$

We need to expand $\Delta F\left(K-k\right) - \Delta F\left(K\right)$. Based on equation (19), we can write

$$
\begin{aligned}
\lim_{N\to\infty}\left(\Delta F\left(K-k\right) - \Delta F\left(K\right)\right) &= \log\frac{P-\left(K-k\right)}{P-\left(K-k\right)+1} - \log\frac{P-K}{P-K+1} \\
&= \log\left(\frac{P-\left(K-k\right)}{P-\left(K-k\right)+1} \times \frac{P-K+1}{P-K}\right), \forall 1 \le k \le K-1.
\end{aligned}
\tag{33}
$$

It is known that $\left(\left(P-K+1\right)\left(P-\left(K-k\right)\right)\right)\big/\left(\left(P-K\right)\left(P-\left(K-k\right)+1\right)\right)$ is always greater than 1 for $1 \le k \le K-1$ and then $\Delta F\left(K-k\right) - \Delta F\left(K\right) > 0$ and based on our assumptions, when *N* tends to infinity

$$
\lim_{N\to\infty} P_{missed} = 0.
\tag{34}
$$

∎

**Note 2**: For criterion given in (21), missed detection occurs when $\Delta F\left(K-k\right) > \Delta F\left(K+l\right)$, for $1 \le k \le K-1$, and $0 \le l \le P-K$. For the computation of the missed detection probability, we should obey a same procedure as it is mentioned for (20).

## V.  SIMULATION RESULTS

In this section we study the performance of the proposed algorithms via computer simulations. Consider a linear array of sensors with complex-valued observations and uncorrelated sources. We compare the performance of the proposed method, i.e. EEE, with MDL, AIC and the methods developed in [16] and [21], denoted by "RMT" and "GBIC$_1$", respectively. The false alarm of RMT, as it is recommended by the authors, is set to 0.1%. We do this by running 10000 independent trials for each point.

Also for the simulations in the presence of non-Gaussian noise we use the Gaussian mixture noise with the probability density function (pdf) given by [38]

$$p(x) = (1-\varepsilon)\mathcal{N}(0,\sigma^2) + \varepsilon\mathcal{N}(0,\eta\sigma^2) \tag{35}$$

where $\varepsilon$ is the probability of impulses in noise and $\mathcal{N}(0,\sigma^2)$ and $\mathcal{N}(0,\eta\sigma^2)$ denote zero mean Gaussian pdf's with variances $\sigma^2$ and $\eta\sigma^2$ respectively. The above model, $p(x)$, is an approximation of the Middleton (Class A) model for impulsive noise [38].

The kernel function that is used for simulations, is Gaussian kernel. The Gaussian kernel is often used because of its convenient mathematical properties [45]. For Gaussian kernel we have

$$\mathcal{K}_G(x) = \frac{1}{\sqrt{2\pi}}e^{-\frac{x^2}{2}}, \quad \mathcal{K}_{G,h}(x) = \frac{1}{\sqrt{2\pi}h}e^{-\frac{1}{2}\left(\frac{x}{h}\right)^2}. \tag{36}$$

An optimum value of smoothing parameter or bandwidth, i.e. $h$ for Gaussian Kernel has been studied in [44], as following

$$h = 1.06\frac{\hat{\sigma}_\lambda}{P^5} \tag{37}$$

where $\hat{\sigma}_\lambda$ is the estimated variance of the $\lambda$ variable. We will use the same value in non-Gaussian noise as well.

The first simulation is shown in Fig. 1, and we examine the probability of detection with respect to the number of snapshots in Gaussian noise by assuming *SNR*=8*dB*, *P*=10, and *K*=5. The probability of detection is considered as the number of runs with true estimated number of sources normalized by the number of runs. It shows that EEE outperforms other methods when the number of observations increases, especially for $N \geq 10$.

The second simulation that is depicted in Fig. 2, is the probability of false alarm versus number of snapshots in Gaussian noise by assuming *SNR*=8*dB*, *P*=10, and *K*=5. In Fig. 3, the probability of missed detection has been shown. This figure shows that, when the number of snapshots increases the probability of missed detection tends to zero. Notice that, the major error for AIC, MDL and GBIC$_1$ is from the false alarm and their probability of missed detection are negligible.

In the Fig. 4, detection probability is shown as a function of SNR in the presence of Gaussian noise. As we see in this figure, EEE has an obvious advantage in performance compared other methods. Our observations shows that the EEE algorithm has a good performance in a few observation samples. This makes that EEE has low complexity of computations in defined performance with comparison of other algorithms.

Fig. 5, shows the probability of detection when the number of sources changes. It is seen that all other algorithms lose their performance when the number of sources increases. The only algorithm that has a negligible degradation in performance is EEE.

In the simulations by assuming the non-Gaussian noise, in Fig. 6 we compare the probability of detection for all algorithms (AIC, MDL, RMT, GBIC$_1$ and EEE) as a function of *SNR* for *N*=100 and non-Gaussian noise parameters $\varepsilon = 0.01$, $\eta = 100$ and $\eta = 1000$, respectively. This figure shows that EEE method is the only algorithm that has acceptable performance in impulsive or non-Gaussian noise.

In Fig. 7 and Fig. 8, the performance of the estimators in various degrees of noise impulsiveness has been shown. They show that EEE has the best performance among the studied methods. We bring the attentions on that, the performance of other algorithms degrade by increasing the impulsiveness of noise. Note that by increasing the probability of impulses, i.e. $\varepsilon \rightarrow 1$, the Gaussian mixture noise goes to the Gaussian model and then, Fig. 8 shows that the performance of other methods than EEE will improve by increasing $\varepsilon$ to more than 0.02. This happens when the performance of EEE is not affected by increasing the impulsiveness degree of noise. In these figures, the detection probability of $GBIC_1$ and MDL is the same as each other.

Fig. 9, shows the probability of detection for different snapshots in non-Gaussian noise. From this figure it can be easily seen that there is a noticeable difference between EEE and other algorithms. In this figure, the $GBIC_1$ and MDL have a same performance.

## VI. Conclusions

In this paper, we present new algorithm based on information theoretic method to determine the number of sources in Gaussian and non-Gaussian noise. We call this algorithm entropy estimation of eigenvalues (EEE). Our approach to entropy estimation is based on kernels and it is shown that we do not need any a priori information of noise. Also we show that the performances of the proposed method outperforms other methods in the literature in Gaussian and non-Gaussian noise.

## Acknowledgment

The authors would like to thank to B. Nadler and S. Kritchman for their RMT [16] and MDL simulation codes.

## Refrences

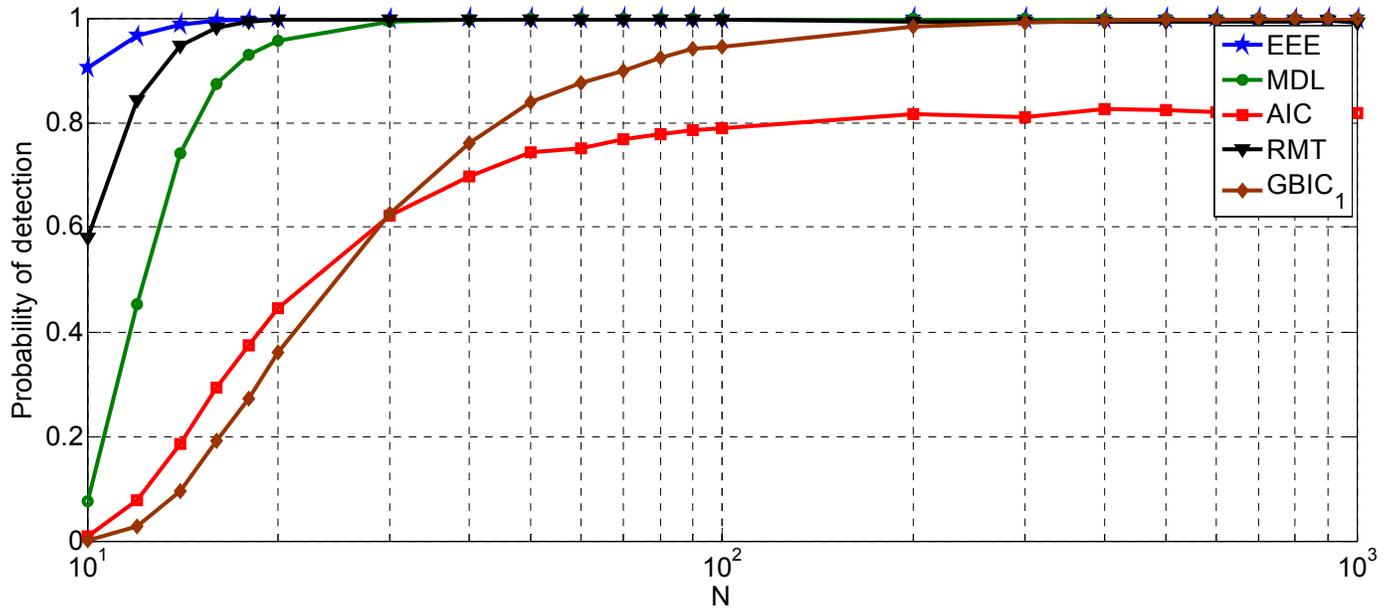

Fig. 1.  Probability of detection as a function of number of snapshots in Gaussian noise: $K = 5$, $P = 10$ and $SNR = 8\ dB$.

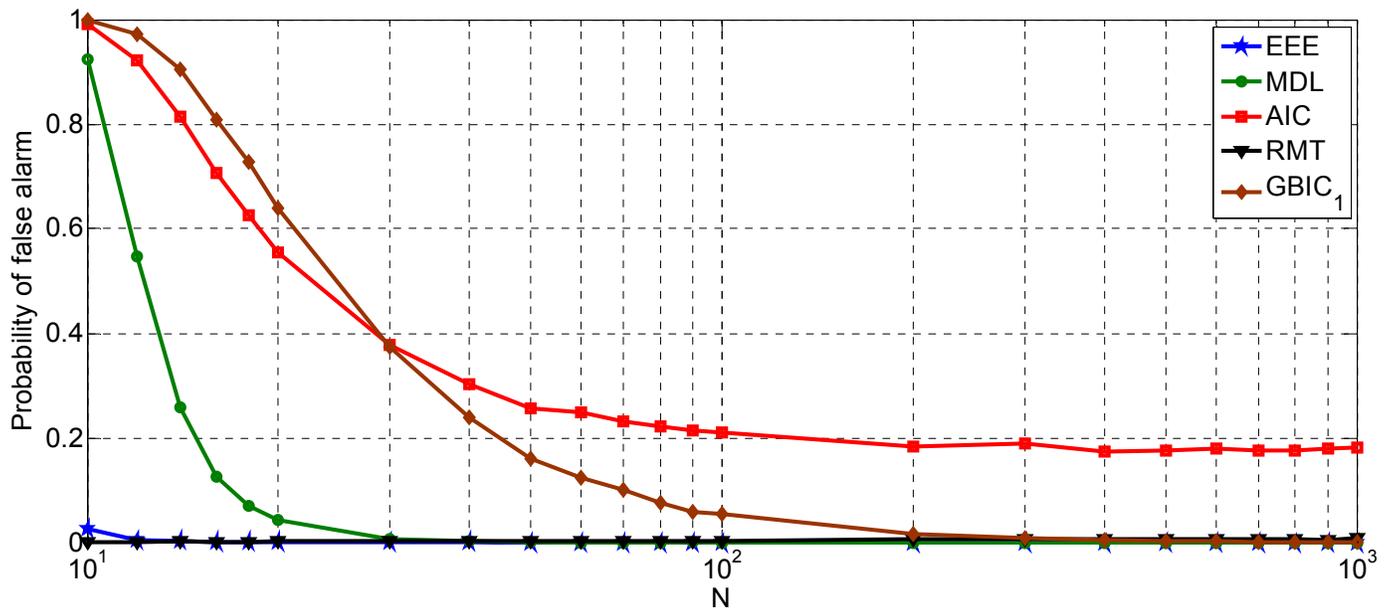

Fig. 2.  Probability of false alarm as a function of number of snapshots in Gaussian noise: $K = 5$, $P = 10$ and $SNR = 8\ dB$.

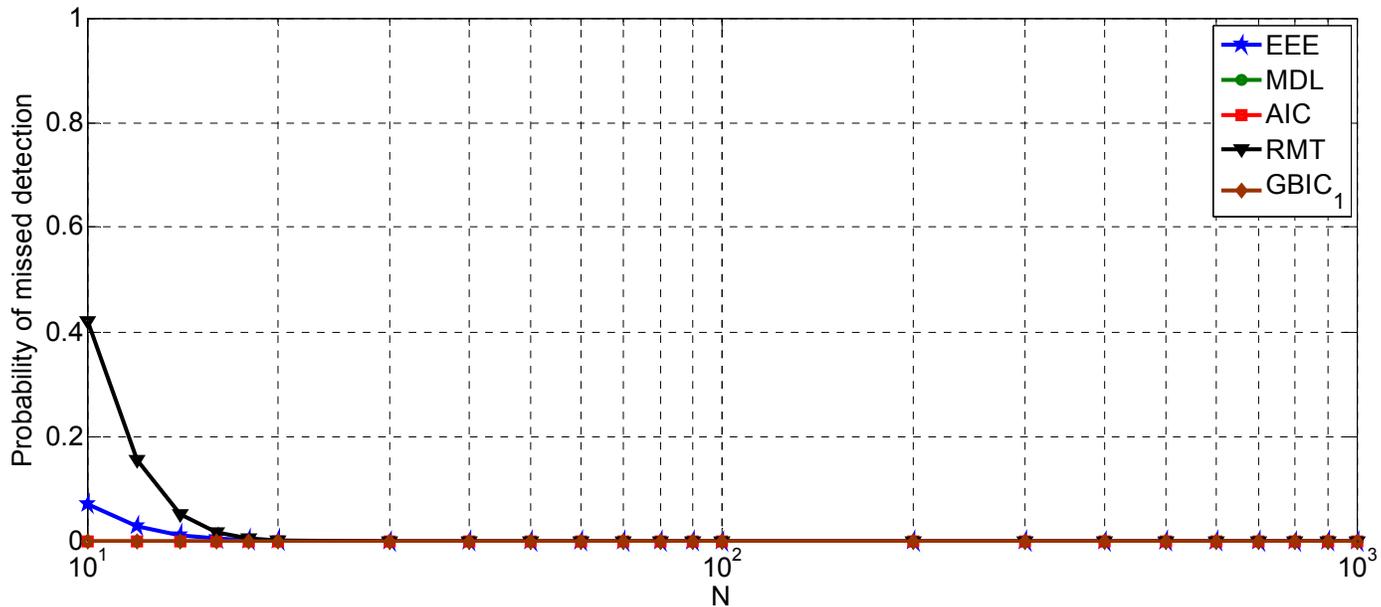

Fig. 3.  Probability of missed detection as a function of number of snapshots in Gaussian noise: $K = 5$, $P = 10$ and $SNR = 8$ $dB$.

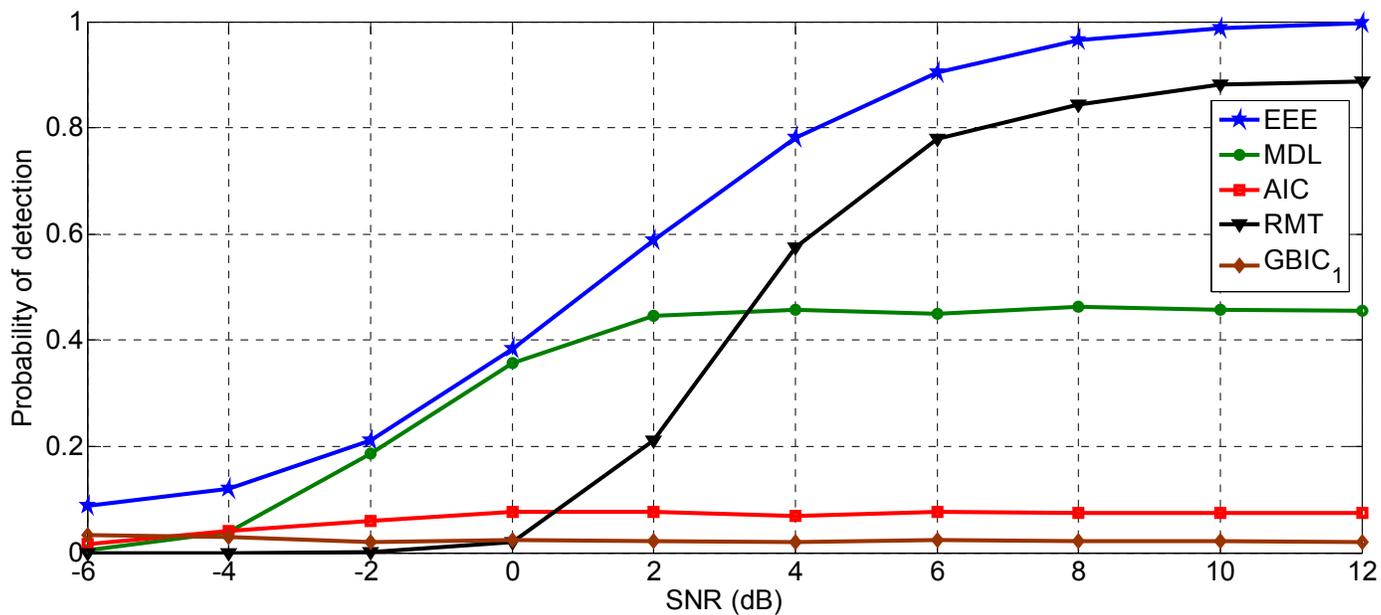

Fig. 4.  Probability of detection as a function of SNR in Gaussian noise: $K = 5$, $P = 10$, and $N = 12$.

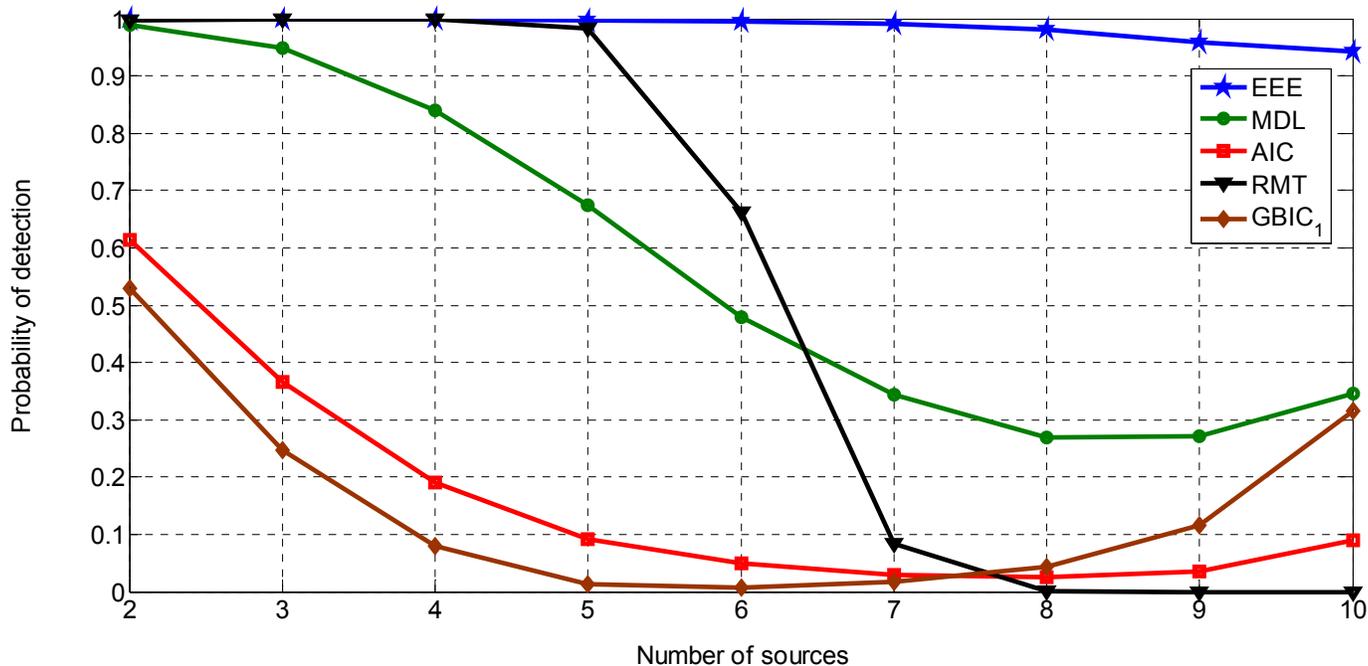

Fig. 5: probability of detection in Gaussian noise as a function of number of sources: $P$=12, $N$=14, $SNR$=8$dB$.

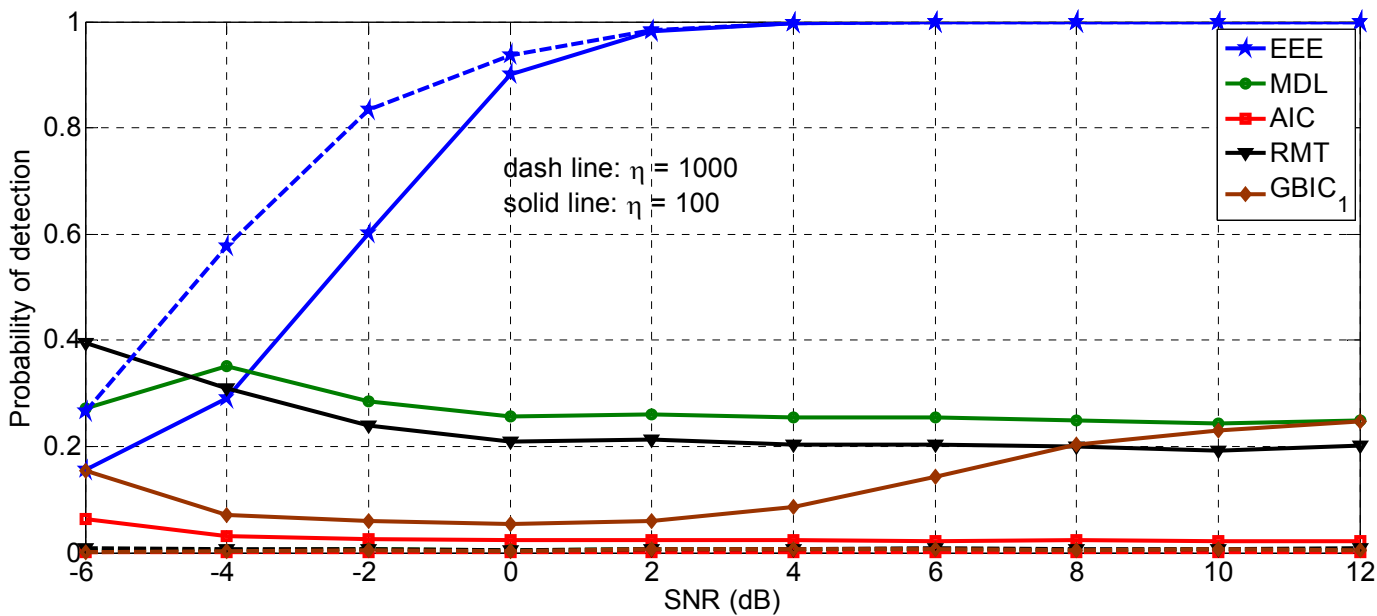

Fig. 6.  Probability of detection in Gaussian mixture noise as a function of SNR: $K$=5, $P$=10, $N$=100, $\varepsilon$=0.01, $\eta$=100, and $\eta$=1000.

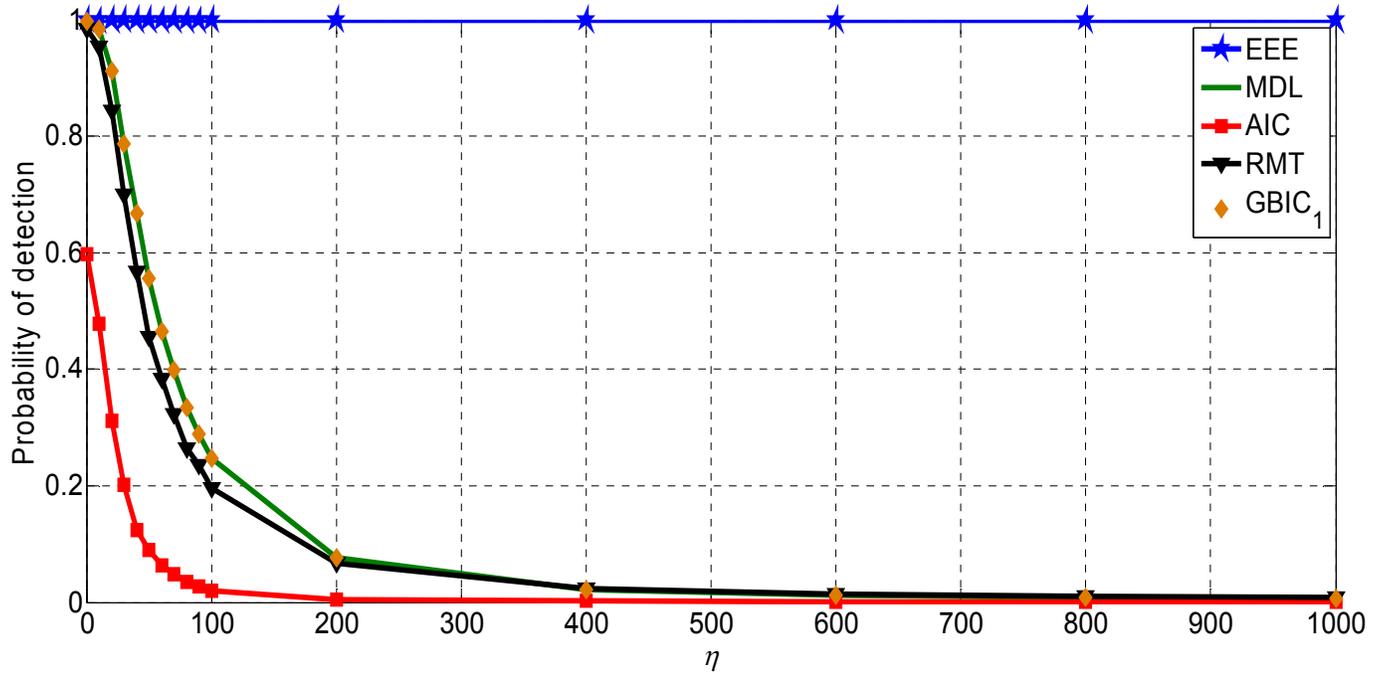

Fig. 7. Probability of detection in Gaussian mixture noise as a function of $\eta$: $K=5$, $P=10$, $N=100$, $\varepsilon=0.01$, $SNR=20dB$.

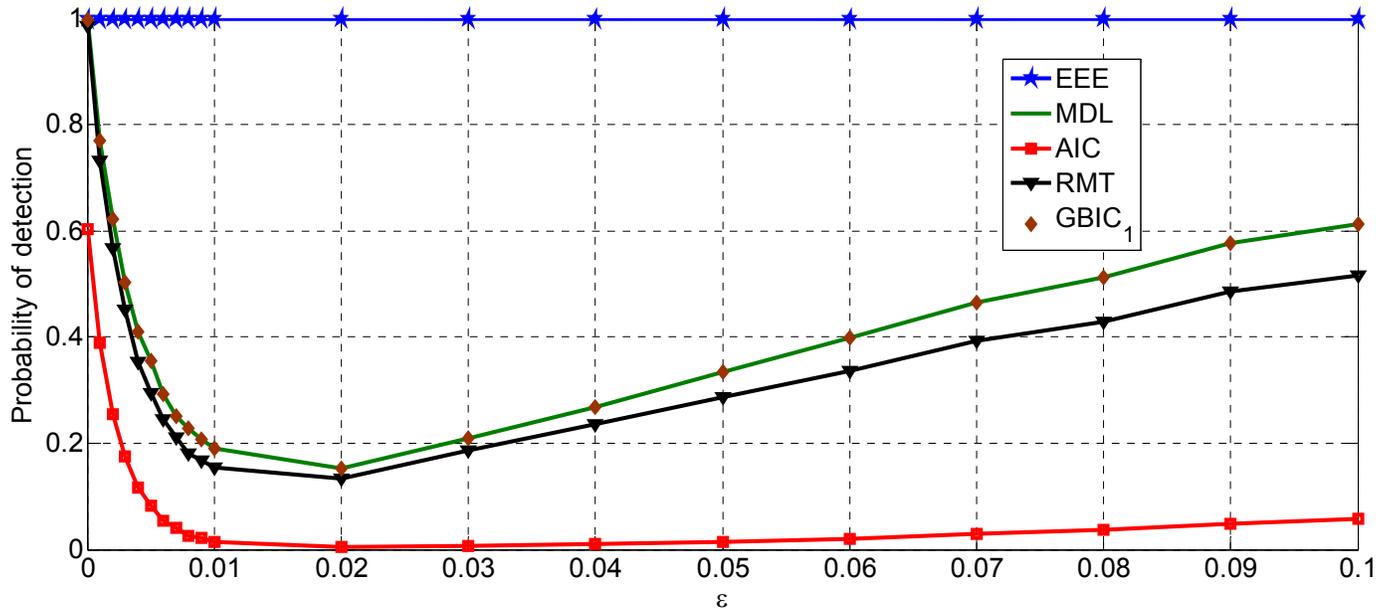

Fig. 8. Probability of detection in Gaussian mixture noise as a function of $\varepsilon$: $K=5$, $P=10$, $N=100$, $\eta=100$, $SNR=20dB$.

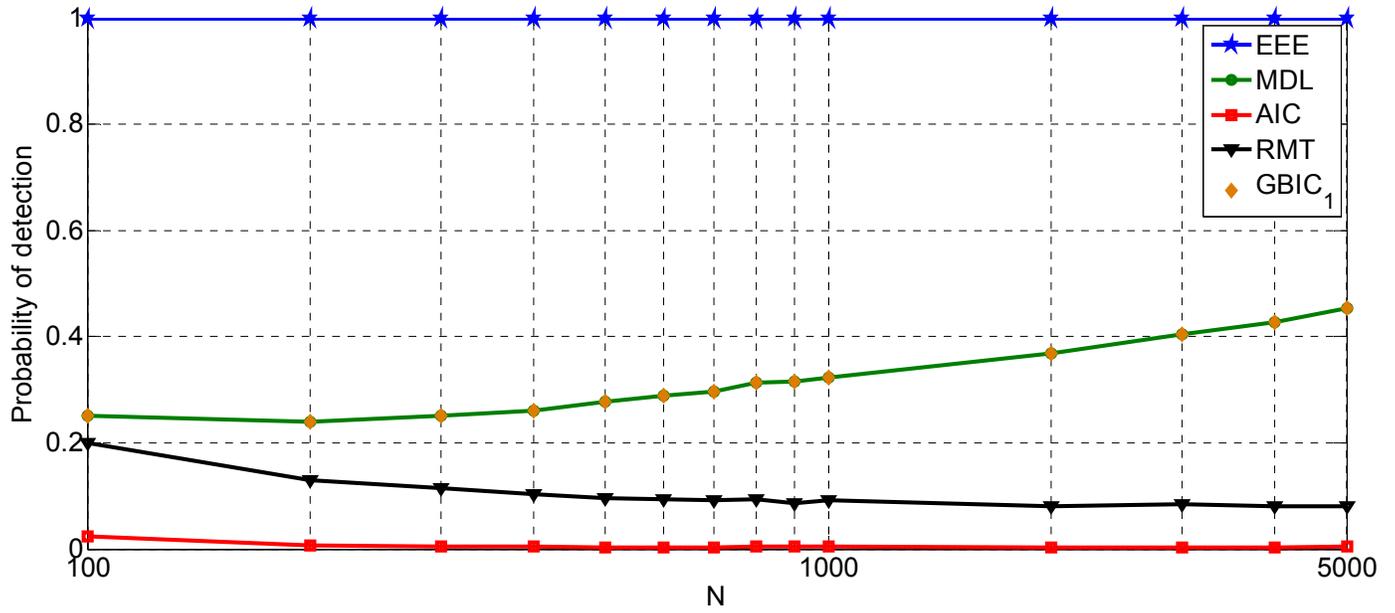

Fig. 9. Probability of detection in Gaussian mixture noise as a function of number of snapshots: $K=5$, $P=10$, $\eta=100$, $\varepsilon=0.01$, $SNR=20dB$.